\documentclass[5p, sort&compress]{elsarticle} 
\usepackage{graphicx}
\usepackage{amssymb}
\usepackage{epstopdf}
\usepackage[amssymb,thinqspace,thinspace]{SIunits}
\usepackage{amsmath}
\usepackage{booktabs}
\usepackage{rotating}
\usepackage{multirow}
\usepackage{psfrag}
\usepackage{adjustbox}
\usepackage{lineno}
\modulolinenumbers[5]
\usepackage{csquotes}
\usepackage{arydshln}
\usepackage{textgreek}
\usepackage{color}
\usepackage{ulem}

\journal{Metall. Mater. Trans. A}

\begin{document}
\begin{frontmatter}
\title{Microstructure evolution and tensile behaviour of a cold rolled 8 wt\% Mn medium manganese steel }
\author[IC]{T. W. J. Kwok}
\author[UOS]{P. Gong}
\author[IC]{X. Xu}
\author[UOS]{J. Nutter}
\author[UOS]{W. M. Rainforth}
\author[IC]{D. Dye\corref{cor1}}\ead{ddye@ic.ac.uk}
\cortext[cor1]{Corresponding author}
\address[IC]{Department of Materials, Royal School of Mines, Imperial College London, Prince Consort Road, London, SW7 2BP, United Kingdom}
\address[UOS]{Department of Materials Science and Engineering, The University of Sheffield, Western Bank, Sheffield, S10 2TN, United Kingdom} 

\begin{abstract}
A novel medium manganese steel with composition Fe-8.3Mn-3.8Al-1.8Si-0.5C-0.06V-0.05Sn was developed and thermomechanically processed through hot rolling and intercritical annealing. The steel possessed a yield strength of 1 GPa, tensile strength of 1.13 GPa and ductility of 41\%. In order to study the effect of cold rolling after intercritical annealing on subsequent tensile properties, the steel was further cold rolled up to 20\% reduction. After cold rolling, it was observed that the strain hardening rate increased continuously with increasing cold rolling reduction but without a significant drop in ductility during subsequent tensile tests. The microstructural evolution with cold rolling reduction was analysed to understand the mechanisms behind this \textcolor{black}{phenomena}. It was found that cold rolling activated additional twinning systems which provided a large number of potent nucleation sites for strain induced martensite to form during subsequent tensile tests \textcolor{black}{in what can be described as an enhanced TRIP effect}. 



\end{abstract}

\end{frontmatter}


\section{Introduction}

Medium Mn steels (4$-$12 wt\% Mn) are an emerging class of duplex ($\gamma + \alpha$) steels that have received considerable research attention. Medium Mn steels have shown to exhibit a successive twinning and transformation induced plasticity (TWIP$+$TRIP) effect if the Stacking Fault Energy (SFE) and stability of the austenite phase are adjusted into the correct regime  during an Intercritical Annealing (IA) heat treatment \cite{Lee2013a,Lee2014}. The TWIP$+$TRIP effect allowed some medium Mn steels to possess large ductilities of 60$-$70\% \cite{Lee2014,Hu2017b,Sohn2014a,Wang2020} and therefore show great potential for use in energy absorbing applications.

Many medium Mn steels which exhibit the TWIP$+$TRIP effect are manufactured by a thermomechanical process composed of hot rolling followed by cold rolling and then IA. The IA is a key step where the solute elements, such as Mn, Al and Si, are able to partition into or out of the austenite phase and therefore alter the SFE and also stability. IA is greatly enhanced with prior cold rolling such that IA can be completed (\textit{i.e.} reaching thermodynamic equilibrium) within several minutes, allowing IA to be conducted on a Continuous Annealing Line (CAL) \cite{Lee2011b} where the heating stage is usually between 2$-$3 minutes long. Otherwise, IA would have to be conducted in batch \textcolor{black}{annealing} furnaces for a minimum of several hours \cite{Savic2018}, severely reducing productivity. 

One potential problem of cold rolling before IA is cracking. The hot rolled microstructure of many medium Mn steels in the literature are either partially or fully martensitic \cite{Sun2017a, Kim2019, Lee2014}. Therefore, a tempering step before cold rolling is usually necessary as the hot rolled strip is potentially very hard with poor ductility \cite{Sun2018}. In conventional strip mills, hot rolled strip is usually 2$-$3 mm thick which then needs to be cold rolled to 1$-$\textcolor{black}{2} mm for automotive applications. The rolling loads necessary to make the reduction in medium Mn steel can be very large as shown in a study by Buchely \textit{et. al.} \cite{Buchely2019}. However, it is anticipated that hot rolling capabilities to produce thinner gauges will continue to improve. For example, recent advances in strip casting technology have shown that medium Mn steel can be successfully cast into 2.5 mm thick strip \cite{Wang2019b}. With this in mind, reliance on cold rolling for heavy gauge reduction can be reduced. Cold rolling can then be pushed to the end of the processing route (\textit{i.e.} hot rolling$-$IA$-$cold rolling) where it is still useful for other purposes such as dimensional control. From an alloy design perspective, element partitioning will no longer be able to rely on cold rolling to improve partitioning kinetics. Ideally, the steel should still be able to exhibit the TWIP$+$TRIP effect but through a simplified hot rolling and IA thermomechanical process route without the need for cold rolling. Several medium Mn steels in the literature have successfully demonstrated this \cite{Lee2015,Lee2015c,Kim2019} but require IA durations between 10$-$30 min which may be challenging in a CAL.

Studies on IA then cold rolled medium Mn steels are relatively sparse compared to studies on cold rolled then IA steels. Sun \textit{et. al.} \cite{Sun2017} studied the cold rollability of high Si, $\delta$-ferrite containing medium Mn steels and found that brittle $\delta$-ferrite could lead to cracking during cold rolling but more importantly that the TWIP$+$TRIP effect greatly enhanced the amount of cold reduction without cracking as compared to another steel which only showed the TRIP effect. Wang \textit{et. al.} \cite{Wang2020} and Li \textit{et. al.} \cite{Li2017a} studied the effect of uniaxial pre-straining on the formation of L{\"u}ders type bands and concluded that pre-straining increases the dislocation density in the austenite, therefore delaying the onset of the TRIP effect which is known to cause the formation of L{\"u}ders type bands \cite{Zhang2017b}. Nevertheless the effect of cold rolling on subsequent tensile behaviour and how it may affect the TWIP$+$TRIP effect is relatively unknown.

This study therefore aims to accomplish two points. First\textcolor{black}{ly} to develop a cost effective \textcolor{black}{(in terms of alloying cost and a simplified processing route)} high strength medium Mn steel that exhibits the TWIP$+$TRIP effect through hot rolling and IA only. Secondly, to investigate the effects of cold rolling (up to 20\% reduction) after IA on tensile behaviour.

\section{Experimental}

In order to minimise cost due to microalloying, a carbide optimisation study was first conducted. Three steels \textcolor{black}{with different V content}  were arc melted from pure elements to produce a 400 g ingot with dimensions of 60 $\times$ 23 $\times$ 23 mm. The \textcolor{black}{nominal composition of Fe-8.3Mn-3.8Al-1.8Si-0.5C-0.05Sn-(0, 0.04, 0.06)V in mass percent} was based on a previous alloy developed by Lee and DeCooman \cite{Lee2015c}. The ingot was sectioned into 4 bars measuring 60 $\times$ 10 $\times$ 10 mm. The bars were quartz encapsulated in low pressure Ar and homogenised at 1250 \degree C for 24 h before quenching in water. Each bar was reheated to 1000 \degree C for 30 min prior to hot rolling. Thermomechanical processing was conducted in 5 passes with decreasing reductions from 1000 \degree C to 850 \degree C with a total reduction of approximately 85\%. The hot rolled strips were quenched immediately after the final pass. After hot rolling, the strips were intercritically annealed at 750 \degree C for 5 min before being allowed to air cool. The 0.06V steel was subsequently cold rolled to 10\% and 20\% reduction. 

Tensile samples with gauge dimensions of 19 $\times$ 1.5 $\times$ 1 mm were machined \textit{via} Electric Discharge Machining (EDM) from the rolled strips such that the rolling direction was parallel to the tensile direction. Tensile testing was conducted at a nominal strain rate of $10^{-3}$ s$^{-1}$ with an extensometer which was removed at 10\% engineering strain. \textcolor{black}{An extensometer was used to accurately measure the Young's modulus and early yielding behaviour but had to be removed at 10\% engineering strain to prevent over-extension of the extensometer.}

Secondary Electron Microscopy (SEM), Electron Backscatter Diffraction (EBSD) and Energy Dispersive Spectroscopy (EDS) were conducted on a Zeiss Sigma FE-SEM equipped with a Bruker EBSD detector and Bruker XFlash 6160 EDS detector. Transmission Electron Microscopy (TEM) was conducted on a JEOL JEM-F200 operated at an accelerating voltage of 200 kV. Phase and orientation mapping was carried out using the NanoMEGAS ASTAR system, which uses Precession Electron Diffraction (PED) to capture EBSD-like maps in the TEM. PED patterns were collected with a precession angle of 0.7\degree, a precession frequency of 100 Hz and a step size of 2.5 nm. The TEM was operated with a spot size of 7 and a 10 \textmu m condenser aperture producing a beam size of minimal diameter ($<$2 nm FWHM). 

Samples for SEM, EBSD and EDS were mechanically ground and polished with an OP-U suspension. Samples for TEM were cut \textit{via} EDM from heat treated blanks, mechanically ground to a thickness below 60 \textmu m and electrolytically polished in a Struers Tenupol twin-jet electropolishing unit using a solution containing 5\% perchloric acid, 35\% butyl-alcohol and 60\% methanol at a temperature of $-40$ \degree C.

\begin{figure}
	\centering
	\includegraphics[width=\linewidth]{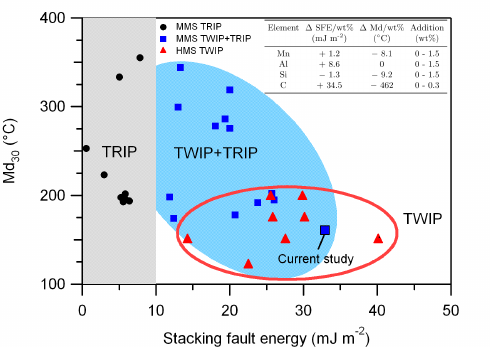}
	\caption{Comparison of SFE and Md\textsubscript{30} of the austenite phase in several Medium Mn steels (MMS) and High Mn TWIP steels (HMS) in the literature. Inset table: Change in SFE and Md$_{30}$ with different additions to austenite with an initial composition of Fe-6.55Mn-1.33Al-1.19Si-0.48C, SFE of 20 mJ m\textsuperscript{-2} and Md\textsubscript{30} of 275 \degree C \cite{Lee2015e}. Data from: \cite{Field2018, Luo2011, Dumay2008, Kim2016b, Lee2015b, Lee2015e, Lee2014, Sun2018}.} 
	\label{fig:sfe-and-md-and-elements2}
\end{figure}


\section{Results}

\subsection{Alloying concept}

There are many considerations to make when designing a medium Mn steel composition and its complementary thermomechanical process route. It is therefore useful to first impose several boundary conditions both in terms of composition and processing. A Mn content of 8$-$8.5 wt\% was chosen as it represented the upper limit of what can be tolerated in conventional secondary steelmelting practice. Si content was chosen to be below 3 wt\% as Si might stabilise and embrittle $\delta$-ferrite at room temperature when added in excess \cite{Sun2019}. Although Al can significantly reduce the density of steel \cite{Sohn2013}, Al content was chosen to be kept below 4$-$5 wt\% as Al is known to clog continuous casting nozzles \cite{Vermeulen2002} and form brittle $\kappa$-carbides in medium Mn steels when added in excessive amounts \cite{Sohn2013,Song2018}. V can also be added for carbide formation and precipitation strengthening \cite{Lee2013b,Lee2015c} but its content should ideally be kept below 0.2 wt\% to minimise the alloying cost of the steel. Lastly, even though Sn is a known tramp element in steel \cite{Zhang2019a}, 0.05 wt\% Sn was deliberately added to potentially improve the galvanisability of the steel \cite{Pourmajidian2019}.

The next step is to consider the composition of the austenite phase. Figure \ref{fig:sfe-and-md-and-elements2} shows a comparison between the stacking fault energy and stability of the austenite phase in several TRIP-type and TWIP$+$TRIP-type medium Mn steels as well as high Mn TWIP steels from the literature. The SFE, if not reported by the authors, was calculated according to the method by Sun \textit{et. al.} \cite{Sun2018}. Austenite stability, represented by Md\textsubscript{30}, is defined as the temperature where half of the total austenite transforms to martensite at a strain of 30\% and can be calculated according to the equation \cite{Angel1954,Nohara1977,Sun2018}:

\begin{equation}
\begin{split}
\small
Md_{30} (\degree C)= \,& 551 - 462 C  - 8.1 Mn  - 9.2 Si ) \\
& - 1.42(-3.29-6.64\log_{10}d-8)
\end{split}
\end{equation}

where compositions are given in wt\% and $d$ is the grain size in \textmu m. A combination of a moderate SFE and low Md$_{30}$ (\textit{i.e.} higher austenite stability) was chosen to promote twinning and a controlled TRIP effect respectively \cite{Herrera2011}. Since the TWIP$+$TRIP mechanism is only operative in the austenite phase, a high austenite fraction is therefore desired ($\geq0.5$) for greater elongation and strain hardening rate.

\begin{table}[h]
	\small
	\centering
	\caption{Bulk composition of 0.06V steel \textcolor{black}{in mass percent} measured by ICP, and IGF for elements marked by $\dagger$.}
	\begin{adjustbox}{width=\columnwidth,center}
		\begin{tabular}{ccccccccc}
			\toprule
			Mn    & Al    & Si    & C$^\dagger$     & V     & Sn    & N$^\dagger$     & P     & S$^\dagger$  \\
			\midrule
			8.29  & 3.78  & 1.81  & 0.467 & 0.06  & 0.047 & 0.001 & $<$0.005 & 0.003 \\
			\bottomrule
		\end{tabular}%
	\end{adjustbox}
	\label{tab:ICPbulkcomp}%
\end{table}%

\begin{figure}
	\centering
	\includegraphics[width=\linewidth]{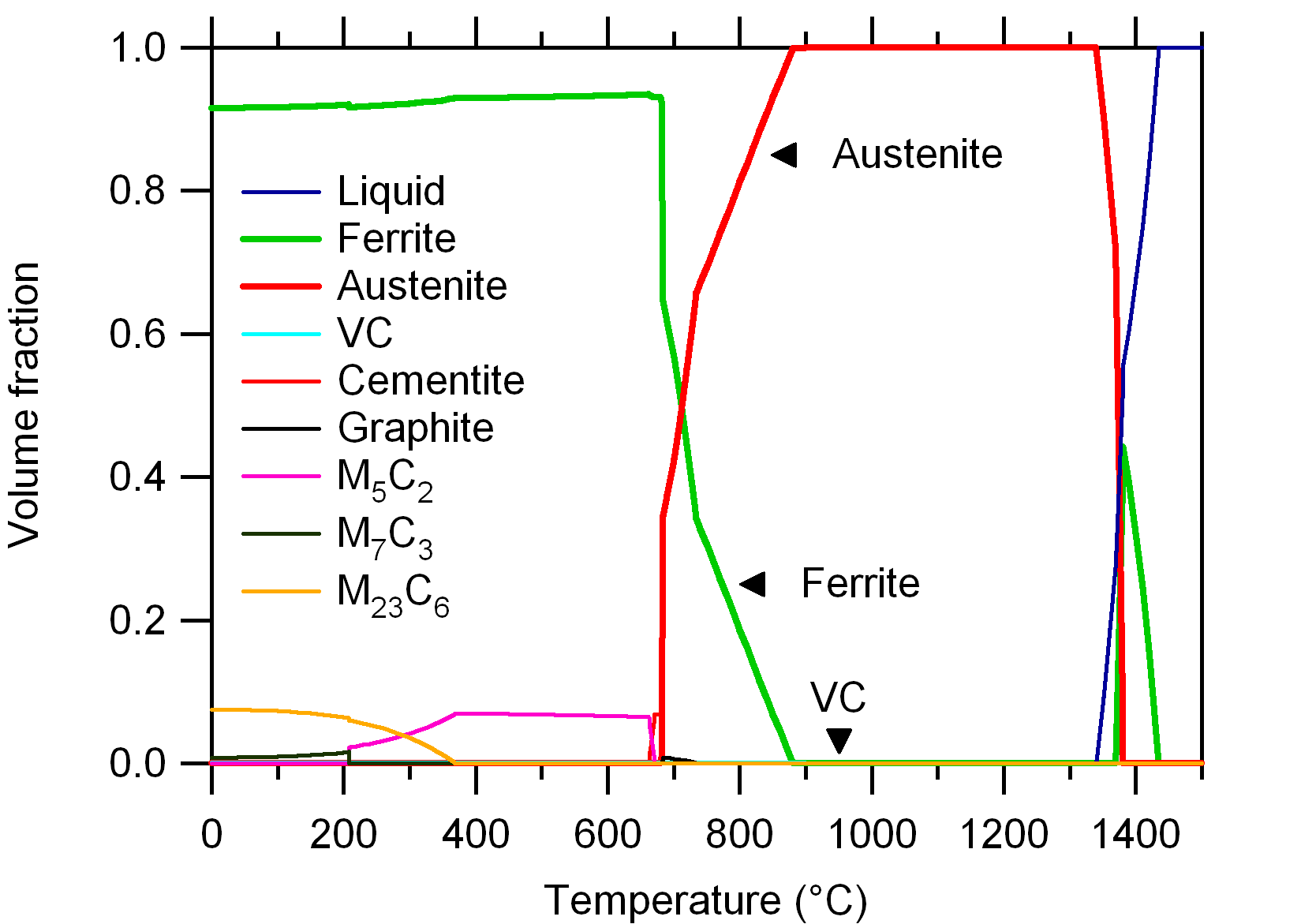}
	\caption{Thermo-Calc property diagram of 0.06V steel.}
	\label{fig:novalloy-thermocalc}
\end{figure}

The thermodynamic software Thermo-Calc is often used to provide guidance on some important processing parameters such as the optimal IA temperature, precipitation temperatures, \textit{etc}. Inductively Coupled Plasma (ICP) and Inert Gas Fusion (IGF) were used to measure the bulk composition of the 0.06V steel and the results are shown in Table \ref{tab:ICPbulkcomp}. The corresponding Thermo-Calc property diagram using the TCFE-7.0 database is shown in Figure \ref{fig:novalloy-thermocalc}. A wide austenitic processing window between 880$-$1200 \degree C was available for hot rolling. The $\delta$-ferrite phase was also not expected to be present at room temperature in this alloy. VC was expected to precipitate at 940 \degree C while graphite and the other M$_x$C$_y$ carbides were expected to precipitate at temperatures 730 \degree C and below. These carbides are thought to be deleterious to ductility and IA should be conducted above 730 \degree C to avoid the formation of these unwanted carbides. Therefore, the practical IA window was between 730$-$880 \degree C where the austenite volume fraction would vary between 0.63$-$1.0 respectively.



\subsection{Tensile properties}

\begin{figure}[t]
	\centering
	\includegraphics[width=\linewidth]{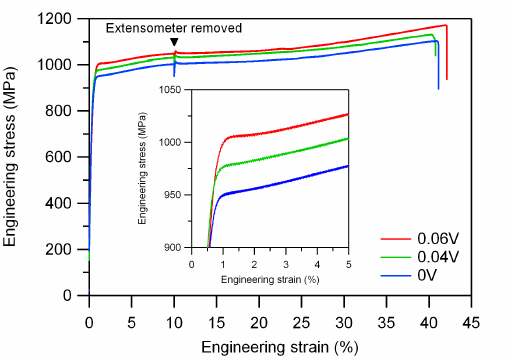}
	\caption{Engineering tensile curves of the investigated steel with different V content. Inset: early yielding behaviour.}
	\label{fig:novalloy-different-v}
\end{figure}

The tensile curves of the investigated steel with different V contents in the as-annealed condition are shown in Figure \ref{fig:novalloy-different-v}. All three steels showed a high yield strength and elongation but a low strain hardening rate. The steels also showed continuous yielding with no Yield Point Elongation (YPE) and the addition of 0.05 wt\% Sn did not appear to deleteriously affect overall tensile properties. 

The addition of up to 0.06 wt\% V resulted in an increase in yield strength from 950 to 1005 MPa but did not affect the strain hardening rate or elongation. Since the addition of 0.06 wt\% V was able to result in a GigaPascal-class steel \cite{Keeler2017}, the 0.06V steel was chosen for further processing.

\begin{figure}[h]
	\centering
	\includegraphics[width=\linewidth]{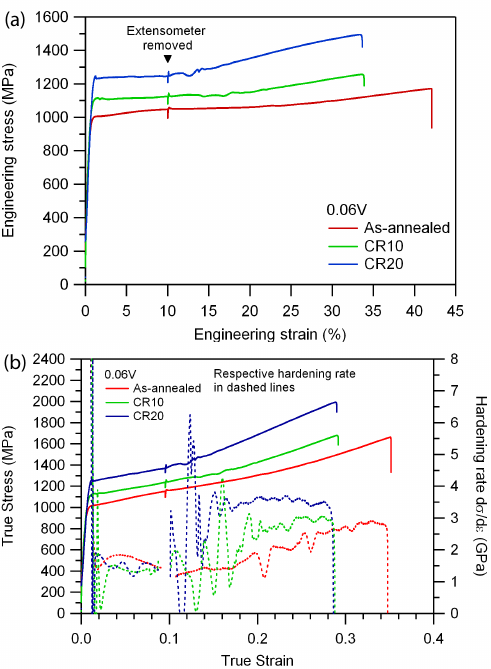}
	\caption{(a) Engineering tensile curves and (b) true tensile curves and hardening rate of the as-annealed and cold rolled 0.06V steel.}
	\label{fig:tensiles-compiled}
\end{figure}

The 0.06V steel was further cold rolled to a reduction of 10\% and 20\% (hereby known as CR10 and CR20 respectively), and the resulting tensile curves are shown in Figure \ref{fig:tensiles-compiled}. In both the CR10 and CR20 samples, a pronounced YPE was observed. The presence of YPE strongly indicates that TRIP rather than TWIP was the dominant deformation mechanism at yielding \textcolor{black}{\cite{Kwok2019,Zhang2017b,Sun2019a,Gao2019}}.

\begin{table}[t]
	\centering
	\caption{Tensile properties of the investigated steels. \textcolor{black}{Note that 0.06V, CR10 and CR20 have the same composition.}}
	\begin{tabular}{lcccc}
		\toprule
		\multicolumn{1}{c}{\multirow{2}[0]{*}{Alloy}} & \multicolumn{1}{c}{$\sigma_{0.2}$} & \multicolumn{1}{c}{$\sigma_{UTS}$} & \multicolumn{1}{c}{$\epsilon$} & \multicolumn{1}{c}{U30} \\
		& \multicolumn{1}{c}{(MPa)} & \multicolumn{1}{c}{(MPa)} & \multicolumn{1}{l}{(\%)} & \multicolumn{1}{l}{(MJ m$^{-3}$)} \\
		\midrule
		0V    & 950   & 1100  & 41    & 361 \\
		0.04V & 985   & 1130  & 41    & 372 \\
		0.06V & 1005  & 1170  & 42    & 379 \\
		CR10  & 1110  & 1260  & 34    & 389 \\
		CR20  & 1250  & 1490  & 34    & 445 \\
		\bottomrule
	\end{tabular}%
	\label{tab:tensileprops}%
\end{table}%

The yield strength, tensile strength and strain hardening rate increased significantly with increasing cold rolling reduction. The strain hardening rate in the as-annealed sample was not constant throughout the tensile curve, most likely due to different deformation mechanisms being active at different stages but averaged out to approximately 2 GPa. In the post-YPE section of the tensile curve, the strain hardening rate of the CR10 sample was approximately 3 GPa. However, in the CR20 sample there was a sustained hardening rate of 3.5 GPa in the post-YPE region. It is also noteworthy that both CR10 and CR20 still retained a large ductility at such high strengths. \textcolor{black}{Another point of interest is the complete lack of post-uniform elongation in all the tensile samples which is also fairly common in other medium Mn steels \cite{Lee2015c,Lee2014,Lee2015e,Hu2017b}.}

A summary of the tensile properties of the investigated steels is shown in Table \ref{tab:tensileprops} including the energy absorption parameter U30 which is determined as the area under the true tensile curve up to a true strain of 0.3 \cite{Kwok2019}.

\subsection{Carbide precipitation microstructure}

\begin{figure}[t]
	\centering
	\includegraphics[width=\linewidth]{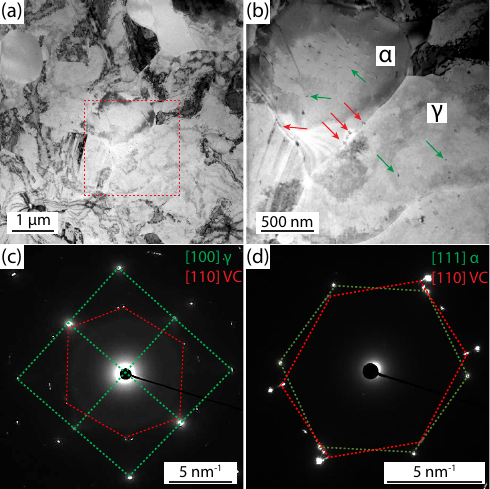}
	\caption{(a) TEM-BF micrograph of 0.06V as-annealed sample. (b) Magnified TEM-BF micrograph of the red square in (a) showing intergranular precipitates (red arrows) and intragranular precipitates (green arrows). Diffraction patterns obtained from (c) an austenite grain, $[100]_\gamma$ zone axis and (d) a ferrite grain, $[111]_\alpha$ zone axis.}
	\label{fig:as-annealed carbides}
\end{figure}

The small strengthening increments shown in Figure \ref{fig:novalloy-different-v} can be attributed to the increase in volume fraction of VC within the microstructure. From the TEM-Bright Field (BF) micrographs in Figure \ref{fig:as-annealed carbides}, small nanometre-sized carbides can be seen in both ferrite and austenite grains and also along grain boundaries. TEM-EDS revealed them to be V-rich in composition, \textcolor{black}{approximately V\textsubscript{4}C\textsubscript{5} but having the cubic structure of VC (Figure \ref{fig:as-annealed carbides}c-d)}. To determine the strengthening contribution of these carbides, the Ashby-Orowan equation after Gladman \cite{Gladman1999} can be used:

\begin{equation}
\Delta\sigma_y^i = \left(\frac{0.538Gbf^{0.5}}{X} \right)\, \ln\left(\frac{X}{2b}\right)
\label{eq:ppt strengthening equation}
\end{equation}

where $\Delta\sigma_y^i$ is the change in yield strength of phase $i$ (MPa), $G$ is the shear modulus ($G_\gamma = 75\,000$ MPa, $G_\alpha = 80\,000$ MPa), $b$ is the Burgers vector ($b_{\gamma} = 0.250$ nm, $b_{\alpha} = 0.248$ nm) \cite{Lee2016}, $f$ is the volume fraction of carbides and $X$ is the mean carbide diameter (nm). The overall increase in strength of the bulk alloy can be described by the rule of mixtures:

\begin{equation}
\Delta\sigma_y^{bulk} = \Delta\sigma_y^{\alpha} \, V_f^\alpha + \Delta\sigma_y^{\gamma} \, V_f^\gamma
\label{eq:ppt strengthening all phases}
\end{equation}

where $V_f^\alpha$ and $V_f^\gamma$ are the volume fractions of ferrite and austentite respectively. Image analysis of TEM micrographs in the 0.06V sample revealed that the average VC diameter was $19 \pm 4$ nm. VC size and distribution appeared to be fairly equal between austentite and ferrite grains. \textcolor{black}{Hu \textit{et. al.} \cite{Hu2019} also observed VC precipitation to be approximately equal in terms of volume fraction and size between austenite and ferrite grains.} The volume fraction of carbides was determined by Thermo-Calc to be $1.25 \times 10^{-3}$. Assuming an equal volume fraction of VC in both austenite and ferrite, the theoretical strength increment from the addition of 0.06 wt\% V as calculated from Equations \ref{eq:ppt strengthening equation} and \ref{eq:ppt strengthening all phases} was 70 MPa. However, the theoretical strength increment was slightly larger than the experimentally observed value of 55 MPa. The difference can be attributed to a slight loss of VC volume fraction to intergranular precipitates at the grain boundaries.

\begin{figure*}[ht!]
	\centering
	\includegraphics[width=\linewidth]{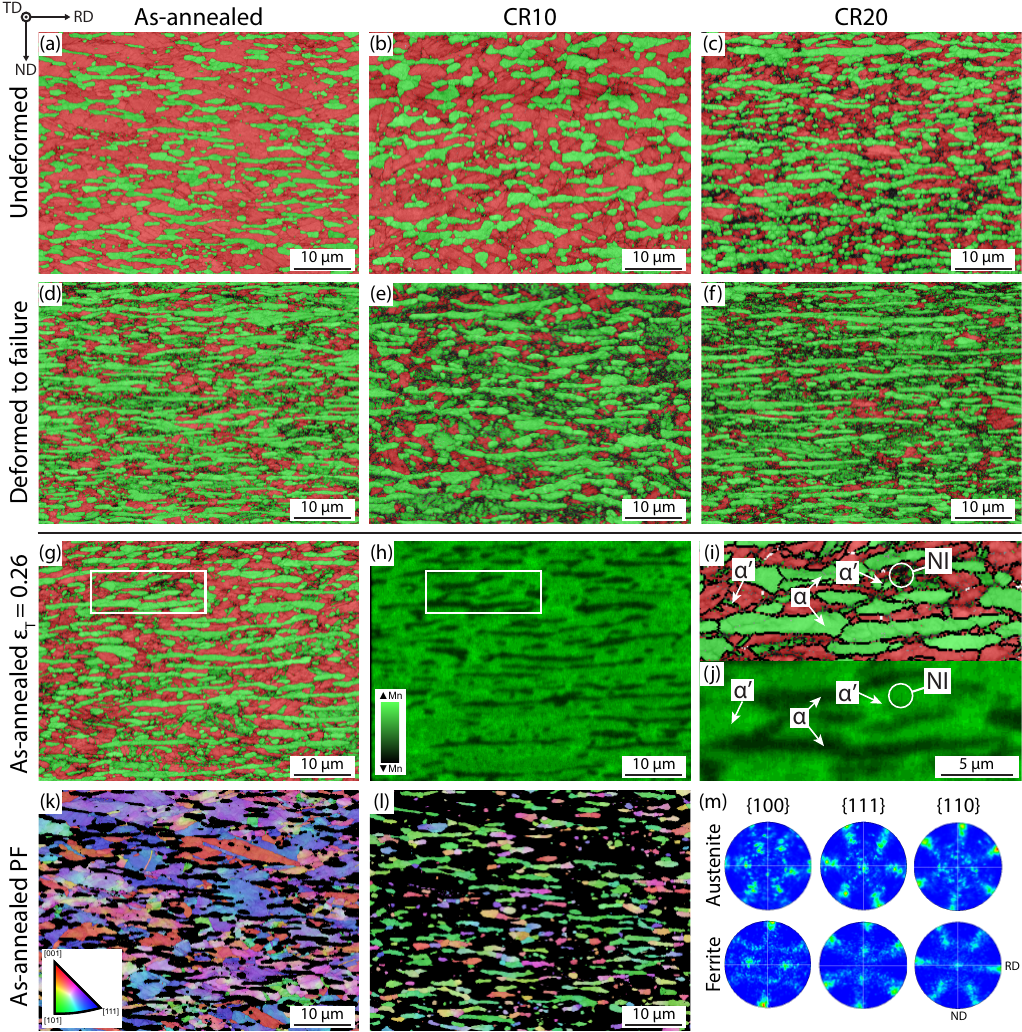}
	\caption{EBSD Image quality (IQ) + Phase Maps (PM) of 0.06V steel in the undeformed (a) as-annealed, (b) CR10, (c) CR20 and deformed to failure (d) as-annealed, (e) CR10, (f) CR20 conditions. EBSD IQ+PM of the (g) as-annealed  0.06V steel deformed to true strain of 0.26, red $-$ austenite, green $-$ ferrite or martensite. (h) Corresponding EDS-Mn map with (h). (i) Magnified region from (g) showing both ferrite and $\alpha$'-martensite. High angle grain boudaries in black and austenite $\Sigma$3 boundaries in white. (j) Corresponding magnified region from (h). (k) Austenite IPF-X, (l) ferrite IPF-X maps and (m) pole figures of the 0.06V as-annealed sample.}
	\label{fig:ebsd-phase-maps}
\end{figure*}

\subsection{EBSD observations}

\begin{figure}[h]
	\centering
	\includegraphics[width=\linewidth]{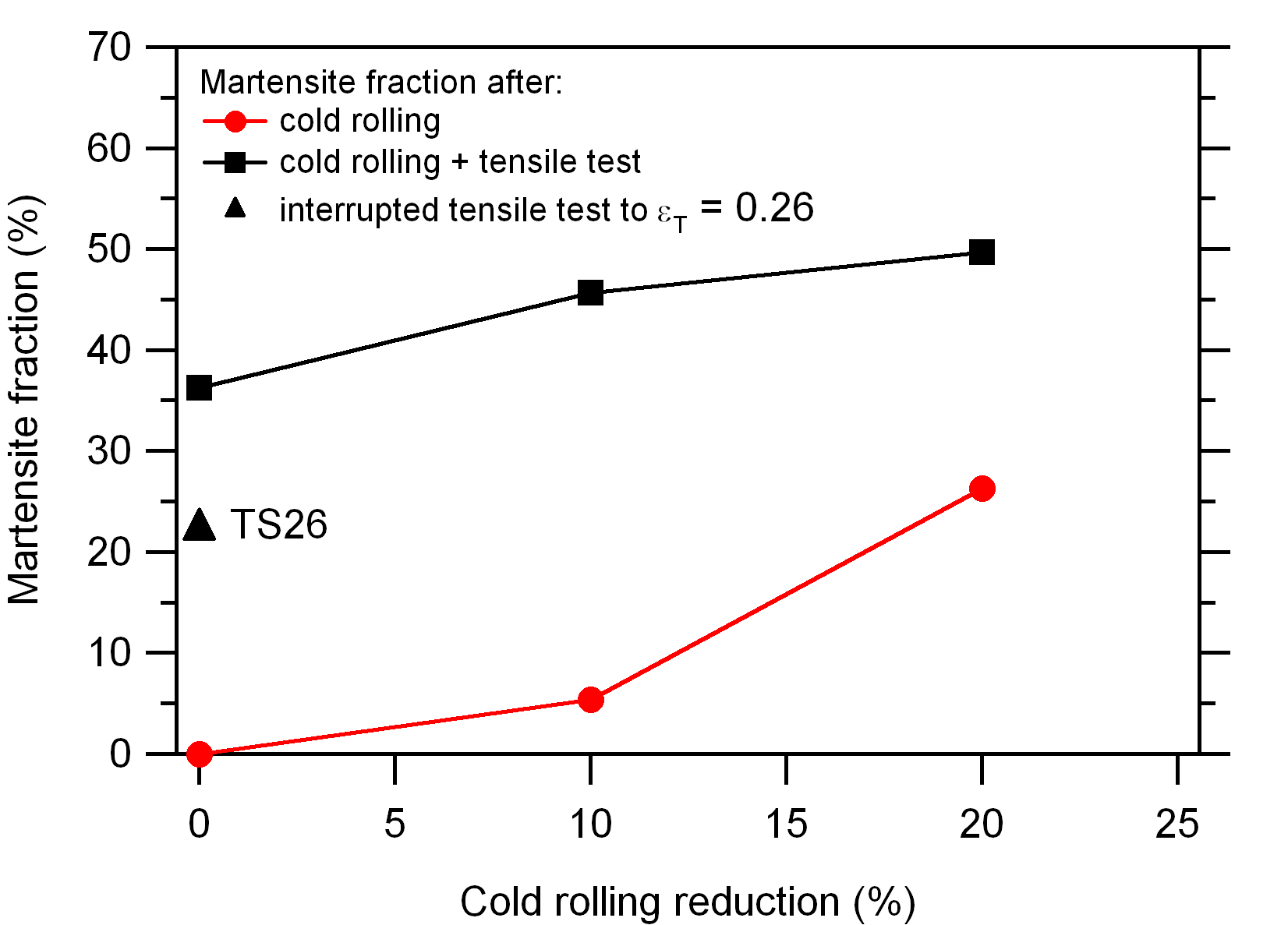}
	\caption{Evolution of martensite fraction with cold rolling reduction before and after a uniaxial tensile test.}
	\label{fig:mart-pct-equivalent-strain}
\end{figure}

The EBSD maps of the as-annealed, CR10 and CR20 0.06V steel are shown in Figure \ref{fig:ebsd-phase-maps} and a summary of the phase fractions is shown in Table \ref{tab:EBSDphasefracs} and Figure \ref{fig:mart-pct-equivalent-strain}. The microstructure of the as-annealed sample did not resemble the typical lath-type microstructure in hot rolled, quenched and IA medium Mn steels \cite{Lee2015c}. Instead, the microstructure can be described as pancaked austenite grains with larger ferrite grains on the austenite grain boundaries and finer ferrite grains in the austenite grain interior. The phase fractions obtained were also reasonably close to that obtained by Thermo-Calc (Table \ref{tab:EBSDphasefracs}) which may suggest that the equilibrium phase fraction had been attained. 

The texture of the as-annealed steel is shown in Figure \ref{fig:ebsd-phase-maps}k-m. The steel was highly textured which is typical of steels rolled below the non-recrystallisation temperature ($T_{nr}$). While annealing twins could be found in multiple austenite grains, the slight curvature of the annealing twins and the lack of a recrystallisation texture imply that the as-annealed microstructure was recovered instead of recrystallised and that the annealing twins formed at higher temperatures during hot rolling. 

Due to the difficulty in distinguishing martensite from BCC ferrite in EBSD, the martensite fraction from any particular map was determined by summing the BCC and Non-Indexed (NI) fractions and subtracting 32.5\%, where 32.5\% is the sum of BCC and NI fractions in the undeformed as-annealed sample (Figure \ref{fig:ebsd-phase-maps}a) and taken to be the baseline. From correlative EBSD and EDS-Mn maps in Figures \ref{fig:ebsd-phase-maps}i-j, most NI areas correspond to a region with high Mn concentration, \textit{i.e.} austenite. This suggests that the NI regions were highly deformed but have the same composition as austenite and therefore may be assumed to be martensite which forms upon deformation. Nevertheless, it is acknowledged that there will be some margin of error with this indirect method of measurement.

From Figure \ref{fig:mart-pct-equivalent-strain}, it is evident that the martensite fraction increased in the cold rolled samples and also in subsequent tensile tests. When cold rolled to 10\% reduction (CR10), the martensite fraction increased slightly, indicating that the TRIP effect was not yet significant. However when cold rolled to 20\% reduction (CR20), the martensite fraction increased signifcantly. A tensile sample made from the as-annealed steel was interupted at a true strain of 0.26, hereby known as TS26, such that it had the same equivalent strain ($\epsilon_{eq}$) as the CR20 sample \cite{Nakada2010}. It was found that the TS26 sample had a slightly lower martensite fraction than CR20. When the cold rolled samples were tensile tested to failure, it was also found that the final martensite fraction also increased with prior cold reduction. The increase in final martensite fraction occured most significantly in the first 10\% of cold rolling reduction (CR10) but began to saturate in the next 10\% (CR20).

\subsection{TEM observations}

\begin{figure}[t]
	\centering
	\includegraphics[width=\linewidth]{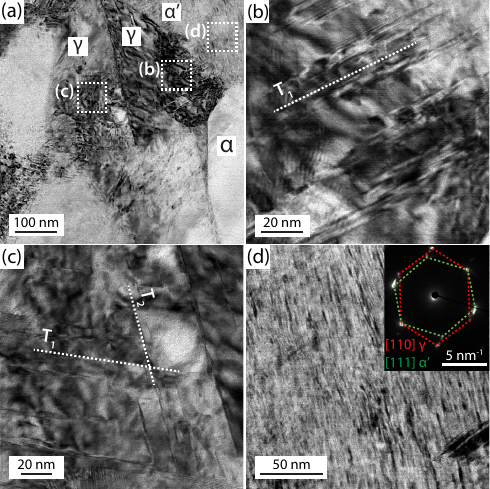}
	\caption{Deformation structures in CR10 sample with beam direction parallel to $[110]_\gamma$. (a) TEM-BF of the general microstructure. (b) STEM-BF of region b showing one active twinning system. (c) STEM- BF of region c showing two twinning systems. (d) STEM-BF of region d showing strain-induced martensite.}	
	\label{fig:cr10-tem}
\end{figure}

While EBSD is helpful in understanding the microstructural changes on a macroscopic scale, it is necessary to probe the deformation structures on a much finer scale using TEM. Figure \ref{fig:cr10-tem} shows the microstructure after 10\% cold rolling reduction, taken from the CR10 sample. The microstructure comprised of austenite, ferrite and also martensite. At a higher magnification, austenite twins were observed in one grain (Figure \ref{fig:cr10-tem}b) and two sets of twins from two twinning systems were observed in an adjacent austenite grain (Figure \ref{fig:cr10-tem}c). The martensite grain in Figure \ref{fig:cr10-tem}d had a diffraction pattern illustrating the K-S orientation relationship with the adjacent austenite grain in region (b). The martensite was therefore likely to be Strain Induced Martensite (SIM) and denoted as blocky-SIM because of its blocky morphology.

\begin{figure*}[t]
	\centering
	\includegraphics[width=\linewidth]{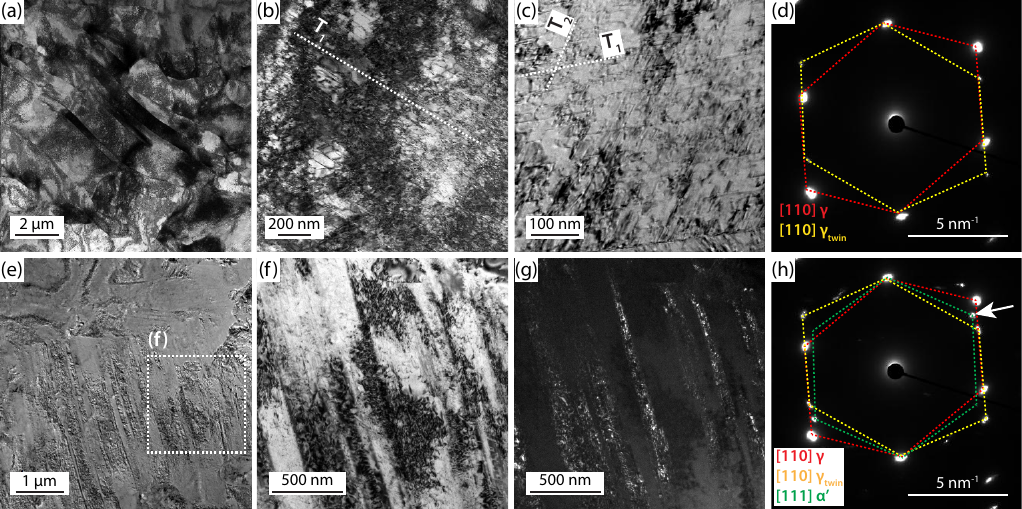}
	\caption{Deformation structures in CR20 sample with beam direction parallel to $[110]_\gamma$. (a) STEM-BF micrograph of the general microstructure. (b) STEM-BF micrograph showing one active twinning system and surrounding high dislocation density. (c) STEM-BF micrograph of a region with two active twinning systems. (d) Diffraction pattern obtained from region in (b). (e) TEM-BF micrograph of an austenite grain showing long lath-structures. (f) Magnified TEM-BF micrograph from (e). (g) TEM-Dark Field (DF) obtained from martensite spot indicated by white arrow in (h). (h) Diffraction pattern obtained from (f).}
	\label{fig:cr20-tem}
\end{figure*}

\begin{figure}
	\centering
	\includegraphics[width=\linewidth]{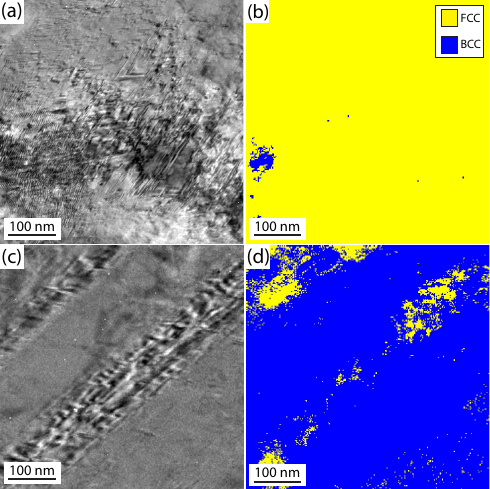}
	\caption{Strain-induced martensite in CR20 sample (a) TEM-BF of two twinning systems and (b) corresponding NanoMegas phase map. (c) TEM-BF of a wide twin and (d) corresponding NanoMegas phase map. Beam direction parallel to $[110]_\gamma$}
	\label{fig:cr20-nanomegas}
\end{figure}

The microstructure after a 20\% cold rolling reduction obtained from the CR20 sample is shown in Figure \ref{fig:cr20-tem}. The microstructure showed that both austenite and martensite phases had a higher dislocation density than the CR10 sample. Austenite grains with a high density of twins from either one twinning system (Figure \ref{fig:cr20-tem}b) or two twinning systems (Figure \ref{fig:cr20-tem}c) were observed. The diffraction pattern in \ref{fig:cr20-tem}d was obtained from the single twinning region in Figure \ref{fig:cr20-tem}b. The NanoMEGAS phase map in Figure \ref{fig:cr20-nanomegas}b revealed some very fine martensite grains in a region similar to Figure \ref{fig:cr20-tem}c containing intersecting twins from two active twinning systems. These martensite grains most likely nucleated at the twin intersections through the mechanism proposed by Olsen and Cohen \cite{Cohen1972,Olson1976b} and is widely reported in TWIP$+$TRIP-type medium Mn steels \cite{Lee2014,Sun2018}.

From Figure \ref{fig:cr20-tem}e, a deformation structure that was not present in the CR10 sample was observed in the CR20 sample. The structure comprised of long wavy laths which spanned the entire grain. Figures \ref{fig:cr20-tem}f-h showed that the wavy laths were most likely kinked austenite twins which contained very fine SIM within the twin boundaries. The SIM was also found to obey the K-S orientation relationship with the parent austenite. However, according to the intersecting shear mechanism by Olson and Cohen \cite{Olson1972}, it would not be possible to nucleate martensite within a single twin. But if the twins were kinked then it might be possible that the deformation may have provided a suitable secondary shear to nucleate martensite within the twin \cite{Bogers1964,Higo1974}. In many Metastable Austenitic Stainless Steels (MASS), martensite can also be found in curved twins without the need for an intersection with another twin \cite{Tsakiris1999,Tian2018,Chen2013c}. MASS are an older class of steels which have been extensively studied. They share a large number of similarities with medium Mn steels such as the TWIP$+$TRIP effect and are therefore a very useful guide to the understanding of medium Mn steels. The SIM which forms at twin intersections and within curved twins shall be denoted as fine-SIM due to its small and particulate morphology.

Another grain in the CR20 sample which also contained similar wavy lath-like structures is shown in Figure \ref{fig:cr20-nanomegas}c-d. However, the NanoMEGAS phase map revealed a different relationship where austenite appears to have nucleated inside a lath within a martensitic matrix. This effect was likely the result of three stages of deformation. Firstly, the formation and thickening of an austenite twin. Secondly, the nucleation of fine-SIM within the twin but not consuming the entire twin. Thirdly, the austenite matrix transforms to martensite, likely due to the growth of blocky-SIM. However, because the austenite phase within the twin was of a different orientation to the austenite matrix, it remains untransformed. This also shows that blocky-SIM grains in the CR20 sample were also growing with increasing cold rolling reduction.

\begin{table}[t]
	\centering
	\small
	\caption{Phase fractions (\%) of the annealed and cold rolled steels before and after tensile testing. $\dagger$ N.I. $-$ Non-indexed percentage. $\ddagger$ Martensite fraction $=$ (BCC + NI) $- 32 - 0.5$.}
	\begin{tabular}{lcccc}
		\toprule
		& \multicolumn{1}{c}{FCC} & \multicolumn{1}{c}{BCC} & \multicolumn{1}{c}{NI$^\dagger$} & $\alpha'\ddagger$ \\
		\midrule
		\smallskip
		Thermo-Calc (750 \degree C) & 71.2 & 28.8 & - & - \\
		0.06V - undeformed & 67.5  & 32    & 0.5 & 0 \\
		0.06V - deformed to $0.26$ & 45.3 & 48.2 & 6.5 & 22.2 \\
		\smallskip
		0.06V - deformed to failure & 31.2  & 59.6  & 9.2 & 36.3 \\
		CR10 - undeformed & 62.1  & 35.2  & 2.7 & 5.4 \\
		\smallskip
		CR10 - deformed to failure & 21.8  & 55.8  & 22.4 & 45.7 \\
		CR20 - undeformed & 41.2  & 47.9  & 10.9 & 26.3 \\
		CR20 - deformed to failure & 17.8  & 61.3  & 20.9 & 49.7 \\
		\bottomrule
	\end{tabular}%
	\label{tab:EBSDphasefracs}%
\end{table}%

\begin{figure*}[t]
	\centering
	\includegraphics[width=\linewidth]{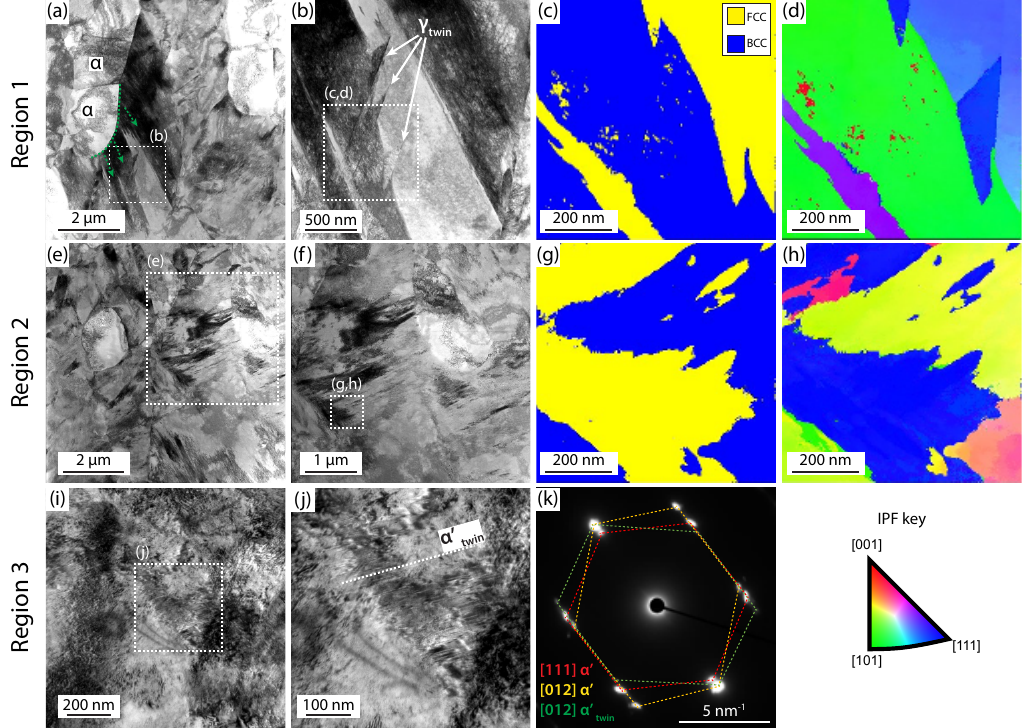}
	\caption{Deformation structures in TS26 sample. From region 1, (a) general microstructure (STEM-BF), (b) magnified view showing martensite and austenite twins (STEM-BF), NanoMegas (c) phase map and (d) IPF from red square in (b). From region 2, (e) general microstructure (STEM-BF), (b) magnified view of austenite and martensite regions (STEM-BF), NanoMegas (g) phase map and (h) IPF from red square in (f). From region 3, TEM-BF micrograph of (i) entirely martensitic region, (j) magnified view showing martensitic twins and (k) diffraction pattern from (j). }
	\label{fig:ts26-tem-nanomegas}
\end{figure*}

The microstructure of the TS26 sample is shown in Figure \ref{fig:ts26-tem-nanomegas} and was very different from the CR20 sample even though they have the same equivalent strain. Three regions from the TS26 sample were chosen in Figure \ref{fig:ts26-tem-nanomegas} to illustrate these differences. In Region 1 (Figure \ref{fig:ts26-tem-nanomegas}a-d), large martensite plates were observed to be propagating outwards in the direction of the dashed green arrows from the interphase grain boundary marked by the dashed green line. The martensite plates appear to be able to propagate through several austenite twins in Figure \ref{fig:ts26-tem-nanomegas}b but with some distortion along the edges of the martensite plate. This suggests that twins are not a strong obstacle to the propagation of blocky-SIM which would render these twins ineffective to the future formation of fine-SIM. Intersecting twins or secondary twins from a second twinning system were not observed in the TS26 sample. 

In Region 2 (Figure \ref{fig:ts26-tem-nanomegas}e-h), the martensite adopted a different morphology with a curved interface as it appears to grow into the surrounding austenite grain. Finally, in Region 3 (Figure \ref{fig:ts26-tem-nanomegas}i-k), twinned martensite or Stress-Assisted Martensite (SAM) was observed. SAM has been shown to be able to form after plastic yielding when the stress is high enough to contribute to a sufficiently large driving force for the martensitic transformation in the absence of alternative nucleation sites for SIM \cite{Olson1972,Maxwell1974,Yen2015}. SAM was therefore likely able to form in TS26 due to a combination of a low density of twins (\textit{i.e.} SIM nucleation sites) and high applied tensile stress (1425 MPa). 


\textcolor{black}{It should be noted that the martensites identified in this steel deviate slightly from the more conventional definitions of SAM and SIM. SAM is usually defined as martensite that forms under applied stress, often below the yield stress of the parent phase and nucleates on pre-existing nucleation sites. SIM is defined as martensite that forms after macroscopic yielding and nucleates on defects generated during plastic deformation \cite{Olson1972,Sinha2003}. As a result, SAM may adopt the morphology of internally twinned plates while SIM tends to form thin laths along slip planes, twins or $\epsilon$-martensite intersections \cite{Maxwell1974}. Here, we have shown that SAM may form after yielding and also the presence of blocky-SIM which nucleates from grain boundaries instead of deformation structures.}

\subsection{Composition}

The composition of each phase of the 0.06V steel was measured by SEM and TEM-EDS and compared against the Thermo-Calc simulation in Table \ref{tab:compositiontable}. The C content in the austenite phase was determined by the lever rule using the phase fractions obtained by EBSD in Table \ref{tab:EBSDphasefracs} and assuming negligible C solubility in the ferrite phase.

Compositions measured by both SEM and TEM-EDS showed that Mn partitioned to austenite, whereas Al and Si partitioned to ferrite, which is well reported in the literature \cite{Hu2017a}. However, the degree of partitioning measured by SEM-EDS was much lower than TEM-EDS. This is largely due to a limitation of the large spot size of SEM-EDS in measuring the composition of a fine-grained material. The TEM-EDS results were therefore taken to be more reliable. While the compositions measured by TEM-EDS do not match up exactly with those predicted by Thermo-Calc, the degree of partitioning was such that the resulting SFE was within the known twinning regime in both high Mn TWIP steels and medium Mn steels. With reference to the Thermo-Calc simulation, the 0.06V steel was at phase equilibrium but not at composition equilibrium, most likely due to the short IA duration.

\section{Discussion}

\subsection{Effect of cold rolling on microstructure}

\begin{table*}[t]
	\centering
	\caption{Composition of the bulk and individual phases in the 0.06V steel measured by ICP, SEM-EDS and TEM-EDS. Uncertainties given in parantheses. $\dagger$C content in SEM and TEM-EDS determined by lever rule based on phase fractions obtained from EBSD. $\ddagger$ Grain sizes calculated from the Bruker ESPRIT software.}
	\begin{tabular}{lccccccccc}
		\toprule
		& \multicolumn{6}{c}{(wt\%)}                    & d\textsubscript{grain}$^\ddagger$ & SFE   & Md\textsubscript{30} \\
		& Mn    & Al    & Si    & C$^\dagger$   & V     & Sn    & (\textmu m) & (mJ m$^{-2}$) & (\degree C) \\
		\midrule
		\smallskip
		Bulk (ICP)  & 8.29  & 3.78  & 1.8   & 0.467 & 0.06  & 0.047 &    -   &   -    & - \\
		
		$\gamma$ (Thermo-Calc) & 9.38 & 3.69 & 1.98 & 0.7  & trace    & -    &  -  & 36.6  & - \\
		$\gamma$ (SEM-EDS) & 8.8 (0.7) & 4.0 (0.6) & 1.9 (0.4) & 0.69  & -    & -    & 4.2   & 38.2  & 165 \\
		\smallskip
		$\gamma$ (TEM-EDS) & 9.5 (0.4) & 3.2 (0.1) & 1.8 (0.1) & 0.69  & 0.07 (0.03) & 0.03 (0.06) & 4.2   & 32.9  & 161 \\
		
		$\alpha$ (Thermo-Calc) & 4.76 & 4.65 & 2.09 & trace      & trace     & -    & -   & -    & - \\
		$\alpha$ (SEM-EDS) & 7.3 (1.2) & 4.2 (0.9) & 1.9 (0.3) & 0     & -    & -    & 1.8   & -    & - \\
		$\alpha$ (TEM-EDS) & 6.5 (0.2) & 4.0 (0.1) & 2.3 (0.1) & 0     & 0.04 (0.02) & 0.04 (0.08) & 1.8   & -    & - \\
		\bottomrule
	\end{tabular}%
	\label{tab:compositiontable}%
\end{table*}%

The TEM observations of the CR10, CR20 and TS26 samples revealed a variety of deformation structures such as twins and martensite in the microstructure. By first comparing the CR20 and TS26 sample, it is evident that the strain path, uniaxial tension in TS26 and plane strain compression in CR10 and CR20, strongly influenced the presence and density of the deformation structures.  The effects of deformation mode on microstructure has been studied extensively in MASS \cite{Nakada2010,Murr1982,Tsakiris1999,Beese2011,Sohrabi2020,Shrinivas1995} and largely show that as the mode of deformation changes from uniaxial to multiaxial, more slip systems and also twinning systems are activated. The same principle also appears to hold true in the investigated steel. While both CR20 and TS26 samples had the same equivalent strain, CR20 had a larger density of twins and even secondary twins (Figure \ref{fig:cr20-tem}) compared to TS26 which only had a low density of primary twins (Figure \ref{fig:ts26-tem-nanomegas}). This shows that cold rolling \textcolor{black}{(\textit{i.e.} plane strain compression)} is able to produce a higher twin density than can be achieved under uniaxial tension. However, the lack of a high density of twins in TS26 was surprising because its SFE as calculated in Table \ref{tab:compositiontable} was well within the known twinning region of TWIP steels \cite{DeCooman2018}. Nevertheless, when compared to other TWIP$+$TRIP-type medium Mn steels in Figure \ref{fig:sfe-and-md-and-elements2}, the investigated steel had the highest SFE. This suggests that twinning kinetics in medium Mn steel are slower and may not be directly comparable with fully austenitic TWIP steels.

Twin intersection density in the microstructure is known to affect the subsequent TRIP effect by providing potent nucleation sites for SIM nucleation \cite{Lee2014,Lee2015b,Lee2016}. Two types of SIM were observed in the investigated steel: fine-SIM and blocky-SIM. Both fine-SIM and blocky-SIM were observed to have nucleated on defects as per the definition by Olson and Cohen \cite{Olson1972} but were distinguished on the basis of morphology and type of nucleating defect. Fine-SIM was only observed in the CR20 sample where it nucleates at twin intersections and in curved twins (Figures \ref{fig:cr20-tem} and \ref{fig:cr20-nanomegas}). Fine-SIM also tends to remain within the twin boundaries even as they grow \cite{Murr1982}.

On the other hand, blocky-SIM was observed to nucleate at grain boundaries, especially interphase grain boundaries (Figures \ref{fig:cr10-tem} and \ref{fig:ts26-tem-nanomegas}) and can grow to very large sizes, even consuming the entire austenite grain (Figure \ref{fig:ebsd-phase-maps}d-f). Blocky-SIM was observed in all three samples CR10, CR20 and TS26 suggesting that the formation of blocky-SIM was unavoidable, regardless of strain path or strain level. However, the existence of blocky-SIM at low strains, \textit{i.e.} CR10 (Figure \ref{fig:cr10-tem}), suggests that grain boundaries, particularly interphase boundaries, can also act as potent nucleation sites for martensite nucleation, allowing martensite to form without the need for twin intersections, contrary to what was suggested by Lee and De Cooman \cite{Lee2015b}. This supports the observation by Sohn \textit{et. al.} \cite{Sohn2014a} that TWIP and TRIP (in the form of blocky-SIM) can occur simultaneously. During the plastic deformation of medium Mn steels, the austenite-ferrite interphase boundary experiences a large amount of strain due to the difference in operating slip systems between the two phases \cite{Lee2014a,Sun2019a}. Therefore, the large strain at the interphase grain boundaries could provide a large driving force for the nucleation of blocky-SIM.

Since twinning and martensite transformation have been shown to occur simultaneously, then TWIP and TRIP can be said to be competitive processes in the investigated alloy. At low strains (\textit{e.g.} CR10) where the plastic mismatch between ferrite and austenite is low \cite{Lee2014a}, only a small fraction of austenite has transformed in the form of blocky-SIM, allowing twins to form in the untransformed austenite. However, at higher strains (\textit{e.g.} CR20 and TS26), a larger fraction of austenite would have transformed to blocky-SIM, with less untransformed austenite available for twinning. Cold rolling therefore promotes twin formation in the austenite before blocky-SIM begins to dominate the microstructure at higher strains, making twinning more difficult.

\begin{figure*}[t]
	\centering
	\includegraphics[width=\linewidth]{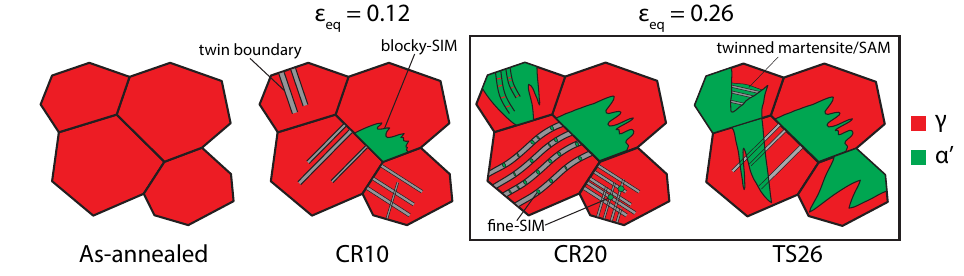}
	\caption{Schematic of the deformation structures in the austenite phase as seen in the various samples. CR10 showing austenite twins, as well as and blocky-SIM. CR20 showing kinked twins with fine-SIM within, intersecting twins and larger-SIM. TS26 showing only one active twinning system, extensive large-SIM and also twinned martensite (SAM).}
	\label{fig:deformation-schematic}
\end{figure*}

A summary of the microstructural evolution of the austenite phase is shown in Figure \ref{fig:deformation-schematic}. At 10\% cold reduction, the austenite phase first shows twins and blocky-SIM. Secondary twins may also be present in certain grains. When cold reduction was increased to 20\%, twinning becomes more extensive and numerous twin intersections have developed. Fine-SIM has started to nucleate in curved twins and at twin intersections while blocky-SIM would also have grown larger and more numerous. Alternatively, at 0.26 true tensile strain, the microstructure contains of a low density of austenite twins, a large fraction of blocky-SIM and also SAM.  

\subsection{Effect of cold rolling on tensile behaviour}

The effect of cold rolling as explained in the previous section was to introduce a variety of deformation structures into the microstructure. When the cold rolled samples were subsequently tensile tested, these deformation structures will certainly be expected to influence the tensile behaviour. The effect of deformation twins is well known and is considered to be beneficial as it strengthens the austenite through the dynamic Hall-Petch effect \cite{DeCooman2018} and also provides nucleation sites for fine-SIM \cite{Lee2014}. Fine SIM, especially twin intersection-type SIM, is also considered beneficial and its effect on tensile properties is well documented and modelled in both medium Mn steels and MASS \cite{Lee2015b,Lee2016,Lebedev2000,Martin2016}. Steels which can form a higher density of twin intersections and therefore form a larger volume fraction of fine-SIM were able to exhibit higher strain hardening rates. However, the effects of blocky-SIM are not as well studied but it is postulated that because of its coarse nature, would not provide as large a benefit to strain hardening or ductility enhancement compared to fine-SIM. SAM which shares a similar blocky and coarse morphology as blocky-SIM is similarly not expected to aid in ductility enhancement \cite{Maxwell1974}.

From Figures \ref{fig:tensiles-compiled} and \ref{fig:ts26-tem-nanomegas}, the as-annealed sample likely deformed \textit{via} simultaneous TWIP and TRIP (blocky-SIM) while also forming SAM at later stages. The lack of fine-SIM and predominance of blocky-SIM in the as-annealed 0.06V steel may explain the lower strain hardening rate as compared to other medium Mn steels in the literature \cite{Lee2014,Lee2015b,Sohn2014a}. On the other hand, the cold rolled samples deformed predominantly \textit{via} TRIP only. The YPE in the cold rolled tensile samples strongly suggests that twinning was no longer active during the early stage of deformation. It was likely that the uniaxial tensile strain was not able to activate new twinning systems nor increase twin density than what was already formed by cold rolling. 

The presence of a high density of potent fine-SIM nucleation sites in the cold rolled samples indicates that during uniaxial tension, the TRIP effect would comprise of the nucleation of both fine-SIM and blocky-SIM. The additional formation of fine-SIM would greatly improve the strain hardening rate and tensile strength which was observed in Figure \ref{fig:tensiles-compiled}. Since the CR20 sample contained a higher twin density and therefore nucleation sites for fine-SIM, the strain hardening rate and tensile strength in the CR20 tensile sample was consequently higher than CR10. The higher martensite fractions in the cold rolled and tensile tested samples shown in Figure \ref{fig:mart-pct-equivalent-strain} may also be attributed to the nucleation and growth of fine-SIM that would otherwise not have formed under pure uniaxial tension. The phenomena of increased martensite fraction in cold rolled then tensile tested samples was also observed in MASS \cite{Milad2008}.

It is also noteworthy that the CR10 and CR20 tensile samples both failed at the same elongation. While the increased final martensite fraction in the cold rolled and tensile tested samples may have contributed to ductility enhancement, Sun \textit{et. al.} \cite{Sun2019} found that in a TWIP$+$TRIP-type medium Mn steel, failure was driven by void formation at $\alpha/\alpha'$ interfaces, most likely due to the large plasticity mismatch between the two phases. \textcolor{black}{Choi \textit{et. al.} \cite{Choi2017} also found that the strain gradient between $\alpha/\alpha'$ interfaces resulted in interface decohesion.} A similar situation may also be occuring in the CR10 and CR20 tensile samples where the interface strength was the limiting factor, since neither sample failed by necking. \textcolor{black}{This failure mode in medium Mn steels may also explain the lack of post-uniform elongation in all the tensile samples in the investigated steel.} 

These findings suggest that the cold rolled samples were no longer TWIP$+$TRIP-type but deformed only \textit{via} the TRIP mechanism. It can be viewed that by cold rolling the steel, the TWIP effect was consumed to produce an enhanced TRIP effect during subsequent uniaxial tensile testing. This implies that the strain hardening rate of medium Mn steels can be varied by simply adjusting the cold rolling reduction after the IA heat treatment. Future alloy design for cold rolled medium Mn steels should aim to reduce the SFE of the austenite to promote twinning during cold rolling and therefore increase the density of nucleation sites for fine-SIM. 

\textcolor{black}{
\subsection{Effect of cold rolling on yield point elongation}
A study by Sun \textit{et. al.} \cite{Sun2019a} explained why medium Mn steels with fine grained equiaxed-type microstructures typically demonstrated considerable YPE. They showed that the large austenite-ferrite interface area, due to the fine grained microstructure, provided a high density of dislocation sources leading to rapid dislocation mutiplication and therefore discontinuous yielding. While the as-annealed microstructure was also fine grained and resembled equiaxed-type microstructures, there was no true recrystallisation during intercritical annealing and it is likely that both austenite and ferrite phases still contained a high density of mobile dislocations after hot rolling, allowing for continuous yielding. }

\textcolor{black}{However, in the cold rolled samples the emergence of the YPE can be attributed to the TRIP effect. Gao \textit{et. al.} \cite{Gao2019} showed how the TRIP effect was able to sustain a large L\"{u}ders strain of 40\% in a metastable austenitic stainless steel SUS304. Since the cold rolled samples contained a high density of SIM nucleation sites, the martensitic transformation would be able to provide fresh mobile dislocations and cause discontinuous yielding. This also explains why the YPE in CR20 was shorter than CR10 as the CR20 sample contains more SIM nucleation sites than CR10 which will result in a higher local strain hardening rate within the L\"{u}ders band and therefore a higher band velocity \cite{Sun2019a}.
}

\subsection{Industrial relevance}

In our previous work \cite{Kwok2019}, we introduced a medium Mn steel of composition Fe-12Mn-4.8Al-2Si-0.32C-0.3V and found that the high price of V greatly increased the alloying cost of the steel. Many researchers have also added V to medium Mn steels and achieved large amounts of precipitation strengthening \cite{Hu2019,Lee2015c}. Hu \textit{et. al.} \cite{Hu2019} was able to improve the yield strength by 650 MPa with 0.7 wt\% V through a warm rolling and IA processing route. However, relying on V for strength is a costly route. In this \textcolor{black}{work}, a carbide optimisation study was conducted and found that 0.06 wt\% V was able to increase the yield strength by 55 MPa and sufficient to raise the yield strength of the as-annealed steel to 1 GPa. 

\textcolor{black}{While the thermomechanical process route was designed to resemble the production of DP steels with a CAL, it is also acknowledged that cold rolling after IA may not be compatible with a Continuous Galvanising Line (CGL) as the final cold rolling step will damage the Zn coating. Nevertheless, as medium Mn steels continue to gain industrial readiness, it is certain that a CGL will eventually be integrated into an optimised process route to produce high strength, high ductility galvanised sheet.}

\begin{figure}[t]
	\centering
	\includegraphics[width=\linewidth]{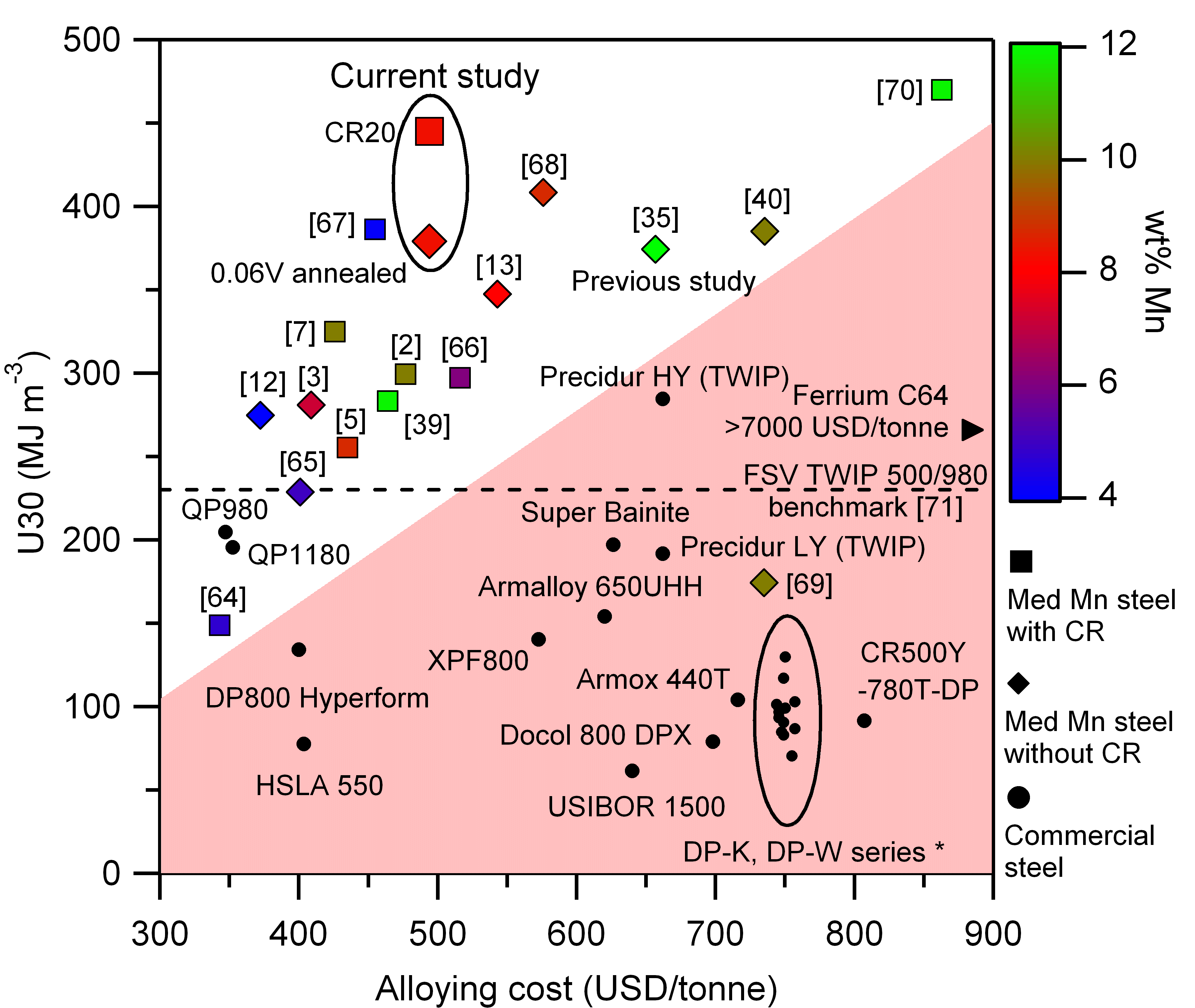}
	\caption{U30 diagram comparing alloying cost and energy absorption between several medium Mn steels in the literature and commerical steels. \cite{Luo2015,Shao2017,Wang2019b,Wang2020, Hu2017b,Lee2016,Lee2014,DeCooman2017,Savic2018,Lee2015,Lee2017,Zhu2017,Kwok2019,Hu2019,He2018a, Sohn2017}. *ThyssenKrupp dual phase steels.}
	\label{fig:novalloy-u30}
\end{figure}

Figure \ref{fig:novalloy-u30} shows an updated figure from our previous work \cite{Kwok2019} which includes the investigated 0.06V steel, several new literature medium Mn steels and also two grades of Quenching and Partitioning (QP) steels from Baosteel. The alloying cost of novel and commercial steels were estimated based on pure metal and ferroalloy additions to steel scrap. \textcolor{black}{It is acknowledged that other processing-related costs and factors also contribute to the total price of steel but it is still helpful for researchers to begin the alloy design process by attempting to minimise alloying cost.} By reducing the Mn and V content from our previous alloy, the alloying cost was significantly reduced by 160 USD/tonne (prices correct as of May 2019) while maintaining a similar U30. The investigated steel in its annealed condition is price competitive with commercially available DP steels yet being two to three times more energy absorbing. When in the cold rolled condition (\textit{i.e.} CR20), the U30 was significantly increased, presenting new opportunities for down gauging in automotive steel. This shows that a holistic alloy design concept can result in a price-competitive high strength and high energy absorbing steel.

\section{Conclusions}
 The effects of cold rolling after IA were investigated in a novel steel and it was found that the tensile properties, especially the strain hardening rate, could be improved with increasing cold reduction. The microstructural evolution with increasing cold rolling reduction was investigated to understand the mechanisms behind the improvement in tensile properties. The key findings from this study are summarised as follows:

\begin{enumerate}
	\item The microstructural evolution of \textcolor{black}{deformation structures in} medium Mn steels \textcolor{black}{are} strain path dependent. Cold rolled specimens showed higher twin densities and active secondary twin systems compared to limited twinning under uniaxial tension without prior cold rolling. 
	
	\item The defects introduced by cold rolling promote the formation of fine-SIM which would lead to an increased strain hardening rate during uniaxial tension, \textit{i.e.} an enhanced TRIP effect.
	
	\item \textcolor{black}{There is strong evidence that both TWIP and TRIP effects were} occuring simultaneously during the deformation of the investigated steel, regardless of strain path.
	
	\item By keeping the V content low \textcolor{black}{($<$0.2 wt\%)}, the investigated steel is price competitive with many commercial steels but with a significantly greater energy absorption capability \textcolor{black}{in terms of crush for automotive crash absorption applications}. 
\end{enumerate}

\section{Acknowledgements}
TWJK would like to acknowledge the provision of a studentship by A*STAR, Singapore. DD, WMR and XX would like to acknowledge funding support from the UK EPSRC (Designing alloys for resource efficiency, EP/L025213/1).

\section{References}
\bibliographystyle{Acta}
\bibliography{library.bib}

\end{document}